\documentclass[12pt, letterpaper]{JHEP3}

\usepackage{amssymb, amsmath, amsopn, amsthm}
\usepackage{epsfig}

\title{\Large BCFW Recursion Relations and String Theory}

\author{Clifford Cheung$^{1,2}$, Donal O'Connell$^3$, and Brian Wecht$^3$ \\
$^1$ Berkeley Center for Theoretical Physics, UC Berkeley,
Berkeley, CA 94720 \\ 
$^2$ Theoretical Physics Group, LBNL,
Berkeley, CA 94720 \\
$^3$ School of Natural Sciences,
Institute for Advanced Study ,
Princeton, NJ 08540\\ 
{\tt clifford.cheung {\rm at}\ berkeley.edu}, {\tt donal {\rm at}\ ias.edu},  {\tt bwecht {\rm at}\ ias.edu}
}

\abstract{We demonstrate that all tree-level string theory amplitudes can be computed using the BCFW recursion relations.  Our proof utilizes the pomeron vertex operator introduced by Brower, Polchinski, Strassler, and Tan. Surprisingly, we find that in a particular large complex momentum limit, the asymptotic expansion of massless string amplitudes is identical in form to that of the corresponding field theory amplitudes. This observation makes manifest the fact that field-theoretic Yang-Mills and graviton amplitudes obey KLT-like relations. Moreover, we conjecture that in this large momentum limit certain string theory and field theory amplitudes are identical, and provide evidence for this conjecture. Additionally, we find a new recursion relation which relates tachyon amplitudes to lower-point tachyon amplitudes.}

\newcommand{\ket}[1]{\mathchoice{
    {\left|{#1}\right\rangle}}{|{#1}\rangle}{|{#1}\rangle}{|{#1}\rangle}}
\newcommand{\sqet}[1]{\mathchoice{
    {\left|{#1}\right]}}{|{#1}]}{|{#1}]}{|{#1}]}}
\newcommand{\bra}[1]{\left\langle{#1}\right|}
\newcommand{\sqra}[1]{\left[{#1}\right|}

\newcommand{\pol}[2]{\epsilon_{#1}^{#2}}
\newcommand{\hpol}[2]{\hat \epsilon_{#1}^{#2}}

\newcommand{\sh}[1]{\langle #1 \rangle}

\newcommand{\half}{\frac{1}{2}}
\newcommand{\quarter}{\frac{1}{4}}

\newcommand{\cA}{\mathcal{A}}

\newcommand{\cN}{\mathcal{N}}
\newcommand{\cO}{\mathcal{O}}

\renewcommand{\t}{\widetilde}

\newcommand{\beq}{\begin{equation}}
\newcommand{\eeq}{\end{equation}}

\newcommand{\h}{\hat}
\newcommand{\apr}{\alpha'}

\def\d{\dot}

\newcommand{\be}{\begin{eqnarray}}
\newcommand{\ee}{\end{eqnarray}}
\newcommand{\MM}{\mathcal{M}}

\DeclareMathOperator{\Tr}{Tr}

\begin{document}
\section{Introduction and Review}
\label{sec:intro}

The last decade has seen remarkable progress in our understanding of scattering amplitudes in quantum field theory. In particular, Witten's twistor string \cite{Witten:2003nn} has inspired a number of novel methods for computing tree-level amplitudes, including the Cachazo-Svrcek-Witten (CSW) rules \cite{Cachazo:2004kj} and the pioneering work of Britto, Cachazo, Feng, and Witten (BCFW) \cite{Britto:2004ap,Britto:2005fq}. The crux of BCFW is that tree-level amplitudes are rational functions of the external momenta---thus, by analytically continuing these momenta into the complex plane, one turns an amplitude into a meromorphic function.  Since a meromorphic function is uniquely determined by its singularities, one can characterize all of the properties of the amplitude by its poles and residues.  The BCFW recursion relations exploit this feature in order to write on-shell amplitudes as sums of products of lower-point on-shell amplitudes.

The validity of the BCFW recursion relations is predicated on the absence of a pole at infinity. This fact has motivated the study of general properties of tree-level amplitudes evaluated at large and complex momenta. Naively, one would expect that tree-level amplitudes could scale with dangerously high powers of $z$, since individual Feynman diagrams contain derivative couplings in both Yang-Mills and gravity.  Remarkably, many amplitudes behave better than expected, and certain amplitudes which  naively blow up at large $z$ actually fall off. In \cite{ArkaniHamed:2008yf}, Arkani-Hamed and Kaplan categorized the helicity-dependent behavior of gauge and gravity amplitudes at large $z$. By interpreting these processes as a hard particle moving through a soft background, they were able to systematically derive the large $z$ behavior of general Yang-Mills and gravity scattering amplitudes.

Despite this progress, the applicability of these new methods to string theoretic amplitudes remains relatively unexplored territory.  To our knowledge, only~\cite{Boels:2008fc} discusses the subject in any detail. The authors of \cite{Boels:2008fc} initiated the study of BCFW techniques in a stringy context by verifying the absence of a pole at infinity for four-point open string gauge boson amplitudes, and by conjecturing that the pole is also absent for higher point amplitudes and for closed string graviton amplitudes. Since string amplitudes often have very good behavior at large momenta, it is reasonable to believe this conjecture. A corollary is that recursion relations along the lines of BCFW should then hold for string theory amplitudes. 

In this work, we address this question in more detail. In particular, we show that all tree-level string amplitudes\footnote{We only discuss tree-level amplitudes involving perturbative string states in this article.} lack a pole at infinity, and so there is a string analog of the BCFW recursion relations. The most important element in our proof is the string pomeron formalism of Brower, Polchinski, Strassler, and Tan \cite{Brower:2006ea}. Using their results, we describe the general large complex momentum behavior of string amplitudes, and show that one can always analytically continue into a region in which these amplitudes vanish at infinity. Additionally, we present an example of stringy recursion relations in the context of bosonic string amplitudes with only external tachyons. In particular, while string BCFW recursion relations necessarily involve an infinite sum over intermediate states of arbitrarily high spins, this sum can be re-expressed in such a way that a tachyon amplitude may be recursively related to lower-point tachyon amplitudes alone.

It is interesting to compare the large $z$ structure of string theory amplitudes with the corresponding series expansion in field theory. For example, one can compare amplitudes involving massless external string states to QFT amplitudes with the same external states. However, we do not expect these series to be simply related because at large $z$ in the string calculation, there is nothing to suppress the effects of the infinite set of massive particles present in string calculations; in other words, it is not clear that the small $\alpha'$ limit commutes with the large $z$ limit. However, we identify a particular limit, which we call the {\it eikonal Regge} limit, of type I string amplitudes. In this limit, massless string amplitudes and their low energy QFT approximation have an identical large $z$ structure. This leads us to suspect that the amplitudes themselves must be related in this limit. We demonstrate that MHV amplitudes in type I string theory and $\cN=4$ super Yang-Mills theory are in fact equal in our limit at four and five points, and provide evidence that a similar simplification occurs at higher points. It is possible that this statement holds more generally, for other string theories or ${\rm N^kMHV}$ amplitudes, but we do not provide evidence for a more general statement. 

It has been observed that asymptotic graviton amplitudes in general relativity exhibit the structure of two copies of  gauge boson amplitudes in field theory. This fact is difficult to explain using purely field theoretic methods, as discussed in~\cite{ArkaniHamed:2008yf}. 
We find that four and five graviton amplitudes in type II string theory are equivalent to general relativity amplitudes in the eikonal Regge limit. If this behaviour continues to hold for an arbitrary number of external gravitons, then the manifest KLT relation in string theory would explain this structure.

The outline of this paper is as follows. In Section 2, we begin by reviewing the BCFW recursion relations, with an emphasis on the relevance of the pole at infinity. We then use the technology of~\cite{Brower:2006ea} to characterize the behavior of stringy amplitudes at infinity and, in particular, to prove that all string amplitudes at tree level can be computed by BCFW recursion. We extend this technology in Section 3 to compute subleading terms in the large $z$ expansion of string amplitudes, and unexpectedly discover a structural similarity between the asymptotic expansion of string amplitudes with massless external states and certain field theory amplitudes. In Section 4 we discuss the relationship of string and field theory amplitudes at large $z$ and our conjecture relating massless string amplitudes to field theory amplitudes in a new limit. We provide evidence that type I string theory and $\cN=4$ super Yang-Mills theory MHV amplitudes are equal in our limit, and show that a conjecture of  Berkovits and Maldacena~\cite{Berkovits:2008ic} implies this correspondence. The subject of Section 5 is how the string theory pomeron expansion can reproduce the field theory asymptotic expansion which was discussed using background field techniques in~\cite{ArkaniHamed:2008yf}. Section 6 focuses on the recursive structure of poles contributing to tachyon amplitudes in bosonic string theory. We conclude in Section 7. Our appendices contain our conventions and some computational details. In Appendix A, we describe our spinor conventions and recall some useful formulae for computing operator product expansions. In Appendix B, we collect some pomeron vertex operators for fermionic states. Finally, in Appendix C, we describe some explicit five-point computations. 

\emph{Note added} : As we completed this work we became aware of Ref.~\cite{Boels:2010bv}. This article has some significant overlap with our work, and in particular also includes a proof of the validity of the BCFW recursion relations in string theory.

\section{Recursions Relations in String Theory}

Let us begin with a review of some of the basic elements of the BCFW recursion relations, so that we can describe how to extend this method to tree-level string theory amplitudes. Our main result will be that string theory amplitudes always vanish at large complex momenta, provided that we work in an appropriate kinematic regime. This fact implies that string amplitudes never have a pole at infinity, and thus obey a version of the BCFW recursion relations. The primary tool we will use to prove this statement is the stringy pomeron developed by Brower, Polchinski, Strassler, and Tan (BPST) \cite{Brower:2006ea}.

\subsection{A Brief Review of BCFW}

The validity of the BCFW recursion relations~\cite{Britto:2004ap,Britto:2005fq} can be understood directly in terms of Feynman diagrams.  In particular, since any tree-level amplitude is built out of propagators and vertices, it must be a rational function of the external momenta.  If one interprets these external momenta as complex variables, then (like any meromorphic function) the amplitude can be reconstructed from its complex singularities.  A key insight provided by BCFW is that the residues at these singularities are equal to products of lower point on-shell amplitudes.

More concretely, BCFW considered on-shell tree amplitudes in Yang-Mills theory\footnote{In this paper we will always take vector boson amplitudes to be color-ordered.} in which the momenta of two external particles are shifted into a light-like complex direction parametrized by a complex number $z$. More generally we can consider such a deformation of an arbitrary amplitude in a quantum field theory. If we take the shifted particles to be particles 1 and 2, this shift is given by
\begin{align}
k_1 \rightarrow \hat k_1(z) &= k_1 + q z \\
k_2 \rightarrow \hat k_2(z) &= k_2 - q z,
\end{align}
where $q$ satisfies $q \cdot q = 0$, $k_1 \cdot q = 0$, and $k_2 \cdot q = 0$.  These constraints are imposed so that the deformed external momenta remain on shell, $\hat k_1^2 = k_1^2$ and $\hat k_2^2 = k_2^2$.  Note also that complex momentum conservation is manifestly preserved.  

One can now show that the deformed amplitude $\mathcal{M}$ is a complex meromorphic function $\mathcal{M}(z)$ that contains only simple poles in $z$ which occur when an intermediate state goes on shell. Notice that we do not need to assume that the propagating particles are massless, as discussed in~\cite{Badger:2005zh}.  It is well-known fact from complex analysis that any meromorphic function which does not have a pole at $z = \infty$ is uniquely determined by its poles and residues at finite $z$.  Assuming $\mathcal{M}(z)$ satisfies this criterion, the amplitude can be written as a sum over its poles at $z = z_k$:
\begin{equation}
\label{eq:bcfwsum}
\mathcal{M}(z) = \sum_k \frac{c_k}{z - z_k}.
\end{equation}
Because each pole in $z$ corresponds to a complex factorization channel, the residues $c_k$ are equal to products of lower point on-shell amplitudes separated by an intermediate on-shell state.  For Yang-Mills theory, the $c_k$ include a sum over the helicity of the intermediate gluon, while more generally there is a sum over all allowed intermediate states.  This construction thus relates on-shell amplitudes to lower-point on-shell amplitudes in a systematic fashion. The absence of a pole at $z=\infty$ is a necessary condition for the BCFW recursion relations\footnote{In the case that there is a pole at infinity, Eq.~\eqref{eq:bcfwsum} still holds, but it must include the pole at infinity.  However, the residue at this pole does not have any physical interpretation in terms of on-shell amplitudes. Methods for dealing with such a pole have been described in~\cite{Benincasa:2007xk} and more recently in~\cite{Feng:2009ei}.}.  Much work has been devoted to extending the original BCFW construction to more general theories in which $\mathcal{M}(z)$ falls off appropriately at large $z$. There now exist recursion relations for massive gauge theories \cite{Badger:2005zh}, gravity \cite{Cachazo:2005ca,Bedford:2005yy}, supersymmetric \cite{ArkaniHamed:2008gz}, and generic field theories \cite{Cheung:2008dn}.

This story translates easily to string theory.  For a string amplitude, an intermediate string propagator can be rewritten as a sum over a ladder of stringy excitations. From the point of view of BCFW, this means that each propagator contributes a variety of simple poles rather than just the gluon.  Therefore, the BCFW recursion relations will also apply to a given string amplitude provided that it falls off appropriately at $z=\infty$. In this section we show that all string amplitudes enjoy a power-law falloff at large $z$ within a particular kinematic regime. This observation is directly related to the celebrated Regge behavior of string amplitudes. Consequently, the BCFW recursion relations can applied to such amplitudes.  Furthermore, since this regime in phase space is an open set, we can analytically continue the resulting recursion relations to string amplitudes at arbitrary kinematic configurations.  

\subsection{BCFW and the Pomeron Vertex}
\label{sec:bcfw}

To derive an analogue of the BCFW relations for strings, we first need to review some technology. Our main tool is the pomeron vertex operator of Brower, Polchinski, Strassler, and Tan (BPST) \cite{Brower:2006ea}. 

We begin with bosonic open string amplitudes before later generalizing. Up to an overall normalization factor, the BCFW-deformed amplitudes of interest are given by
\begin{equation}
\label{eq:stringAmp}
\mathcal{M}(z) =  |w_{N,1} w_{N-1,1} w_{N,N-1}  | 
 \int \!\! \left( \prod_{i = 2}^{N-2} dw_i \right) \! \langle V_1(\hat k_1, w_1) V_2(\hat k_2, w_2)  V_3(k_3, w_3) \cdots V_N(k_N, w_N) \rangle,
\end{equation}
where $w_{i,j} = w_i - w_j$ and $V_i(k_i, w_i)$ is the vertex operator for the $i$th particle with momentum $k_i$ inserted at position $w_i$ on the worldsheet. Note that we have chosen to fix the SL(2, $\mathbb{R}$) invariance by fixing the locations of the vertex operators for particles $1$, $N-1$ and $N$. We have deformed particles 1 and 2 and will be interested in the behavior of this amplitude in the large $z$ limit. We will work in Minkowski spacetime, in which case $z$ must be complex. We could equivalently work with real $z$ in a spacetime with two timelike directions.

We now summarize the result of BPST, whose technology will play a central role in our work.
Consider the product of two open string tachyon vertex operators
\begin{equation}
\int dw \; e^{i \hat k_1 \cdot X(0)} e^{i \hat k_2 \cdot X(w)} = \int dw \; w^{2 \alpha'  k_1 \cdot k_2} e^{i [\hat k_1 \cdot X(0) + \hat k_2 \cdot ( X(0) + w \dot X(0) + \cdots)]}
\label{pomint}
\end{equation}
where on the right-hand side we have performed the OPE, applied a Maclaurin expansion in $w$, and used $\hat k_1 \cdot \hat k_2 = k_1 \cdot k_2$. The key insight of BPST is that when $z$ is large, the integral in Eq.~\eqref{pomint} is dominated by $w \sim 1/z$. This is simply because of the large $z$ in the exponent; unless $w$ is small, large fluctuations in this exponent lead to a negligible contribution to the integral. This small $z$ region corresponds to the limit in which the vertex operators for particles 1 and 2 are close together on the worldsheet.  

An accounting of the $z$ expansion shows that contractions of $ w \; \hat k_2 \cdot \dot X(0)$ into other vertex operators in the string amplitude are of order $1$ while the terms in the ellipsis in Eq.~\eqref{pomint} are of order $1 /z$ or higher, since they come with more powers of $w$. We may thus truncate the expansion and perform the integral, obtaining
\begin{eqnarray}
\int dw \; e^{i \hat k_1 \cdot X(0)} e^{i \hat k_2 \cdot X(w)} &\sim&  \Gamma(-1 - \alpha' s_{12}) [-i \hat k_2 \cdot \dot X(0)]^{1 + \alpha' s_{12}} 
e^{i k \cdot X(0)} ,
\end{eqnarray}
which is the pomeron vertex operator for the tachyon. Here, we have rewritten $k \equiv \hat k_1 + \hat k_2 = k_1 + k_2$. Additionally, we define $s_{ij} \equiv -(k_i + k_j)^2$. 

Had we started with external particles other than tachyons, we would have found a generalized pomeron vertex operator
\begin{eqnarray}
\label{eq:pomVertex}
\int dw \; V_1(\hat k_1, 0)V_2(\hat k_2, w) \sim  C_{12}(z)\Gamma(-1 - \alpha' s_{12}) [-i \hat k_2 \cdot \dot X(0)]^{1 + \alpha' s_{12}} 
e^{i k \cdot X(0)} .
\end{eqnarray}
where $C_{12}(z)$ is a rational function of $z$, and $V_1$ and $V_2$ are the vertex operators corresponding to particles 1 and 2.  In the case where 1 and 2 are tachyons, $C_{12}(z)=1$. Alternatively, if these particles are gauge bosons, then $C_{12}(z)$ arises from contractions with polarization vectors that come with the gauge boson vertex operators,
\begin{equation}
V_i(k_i, w_i) = \epsilon_i \cdot \dot X(w_i) e^{i k_i \cdot X(w_i)},
\end{equation}
where $\epsilon_i$ is the polarization vector of the gauge boson satisfying $\epsilon_i \cdot k_i = 0$. In this case, a simple computation shows that
\begin{equation}
C_{12}(z) = -2 \alpha' (\hat \epsilon_1 \cdot \hat \epsilon_2 - 2 \alpha' \; \hat \epsilon_1 \cdot k \; \hat \epsilon_2 \cdot k),
\label{cgbs}
\end{equation}
where $\hat \epsilon_{1,2}$ are shifted polarizations satisfying $\hat \epsilon_i \cdot \hat k_i =0$. 
The pomeron vertex operator has the key property that it isolates the dependence of the amplitude on $z$ in two terms: a rational function $C_{12}(z)$, and the operator $[-i \hat k_2 \cdot \dot X(0)]^{1+  \alpha' s_{12}}$.   The power $1 + \apr s_{12}$ in Eq.~(\ref{eq:pomVertex}) can be fixed by requiring that the pomeron vertex operator should have dimension one.

It is now straightforward to understand why the BCFW recursion relations can be applied to open bosonic string amplitudes.  From Eq.~\eqref{eq:pomVertex} it is clear that the large $z$ dependence of any bosonic open string amplitude is
\begin{equation}
\mathcal{M}_{\rm open}(z) \sim z^{n + 1 + \alpha' s_{12}}
\end{equation}
where $C_{12}(z)$ grows like $z^n$ at large $z$. If we restrict to a kinematic regime in which $n + 1 + \alpha' s_{12} < 0$,
then there can be no contribution to the amplitude from a pole at infinity, and the BCFW recursion relations can be applied.

In contrast to the usual BCFW procedure, it may seem strange that we need to go to a particular kinematic regime to get the good behavior we want, and then argue by analytic continuation that the amplitude should behave nicely. However, this procedure is common in the context of string amplitudes. Even in the Veneziano amplitude, the integrals one must compute only converge in an unphysical kinematic regime. We can consistently analytically continue these expressions as long as our function is well-defined in a open set. Since our kinematic region is indeed an open set,
we may analytically continue without any problems. The analytic continuation of the amplitude is unique, and so it determines the function in a physical regime. 

It is straightforward to generalize these results to the closed bosonic string. The pomeron vertex operator for closed string states \cite{Brower:2006ea} is
\begin{equation}
\int d^2w \; V_1(\hat k_1, 0)V_2(\hat k_2, w) \sim  C_{12} (z) \Pi (\alpha' s_{12}) e^{i k \cdot X(0)} [ \hat p_2 \cdot \partial X(0) \hat p_2 \cdot \bar \partial X(0)]^{1 + \alpha' s_{12} /4},
\label{clpom}
\end{equation}
where $C_{12}(z)$ is a rational function in $z$ growing like $z^n$ at large $z$, and
\begin{equation}
\Pi (\alpha' s_{12}) = 2 \pi \frac{\Gamma (-1 - \alpha' s_{12}/4)}{\Gamma( 2 + \alpha' s_{12}/4)} e^{- i \pi - i \pi \alpha' s_{12}/4}.
\label{pompi}
\end{equation}
For more details on the derivation of Eq.~\eqref{clpom}, see \cite{Brower:2006ea}.
With this result in hand we can immediately deduce the asymptotic behavior of closed string amplitudes as a function of $z$:
\begin{equation}
\mathcal{M}_\mathrm{closed}(z) \sim z^{n + 2  + \alpha' s_{12}/2}.
\end{equation}
Therefore BCFW recursion relations hold for the closed string in the region $ n + 2 + \alpha' s_{12}/2 < 0$. Any amplitude can be computed recursively in this region and then analytically continued. 

We can easily derive the pomeron vertex operators for superstrings as well. Superstring vertex operators are necessarily written in different pictures, corresponding to whether or not we integrate over their worldsheet superspace coordinates. NS sector operators can be in either the -1 (not integrated) or 0 (integrated) picture, and R sector operators can be in either the -1/2 or +1/2 pictures. As an example, the vertex operators for type I gauge bosons are given by
\begin{eqnarray}
V_{-1}&=&  \epsilon_\mu \psi^\mu  e^{ik \cdot X} e^{-\phi} , \\
V_{0} &=&   \left ( 2\apr \right)^{-1/2} \epsilon_\mu \left ( i \dot X^\mu + 2\apr k \cdot \psi \, \psi^\mu \right ) e^{i k \cdot X}
\label{vops}
\end{eqnarray}
in the $-1$ and $0$ pictures respectively, 
where $\psi^\mu$ is a worldsheet fermion and $\phi$ is a bosonized superconformal ghost. We have not written the gauge group generator, since we are suppressing color structure. The type II graviton vertex operators can be read off from Eq.~\eqref{vops} essentially by taking one copy of the type I vertex operators on each side of the string, and additionally taking $\apr \rightarrow \apr/4$. The heterotic vertex operators are the same as the open string on one side of the string, but include a current $j^A$ on the non-supersymmetric side.

In writing down our pomeron vertex operators, we need to choose the picture of the vertex operators with shifted momenta. Although the eventual amplitude is the same regardless of picture, a convenient choice will make some things easier for us to read off. For the NS sector, the -1 picture is computationally mildly easier to deal with, since the vertex operator has only one term. However, in this section we will work in the 0 picture, since it is in this picture that the physics is most manifest. 
We find that the relevant pomeron vertex operators are
\begin{subequations}
\begin{align}
 {\rm Type\, \, I:} &\,  (\hat \epsilon_1 \cdot \hat \epsilon_2) (1 + \apr s_{12}) \Gamma(-1 - \alpha' s_{12}) [-i \hat k_2 \cdot \dot X(0)]^{1+ \alpha' s_{12}} e^{i k \cdot X(0)} \label{pomsA} \\
  {\rm Heterotic:} &\,(\hat \epsilon_1 \cdot \hat \epsilon_2)\left (1 + \frac{\apr s_{12}}{4} \right ) \Pi (\apr s_{12}) [ \hat k_2 \cdot \partial X(0) \hat k_2 \cdot \bar \partial X(0)]^{1 + \frac{\alpha' s_{12}}4} e^{i k \cdot X(0)} \\
    {\rm Type\,\, II:} &\,  (\hat \epsilon_{1\mu \nu} \hat \epsilon_2^{\mu \nu}) \left  (1 + \frac{\apr s_{12}}{4} \right)^{\!\!2}  \Pi (\apr s_{12})  [ \hat k_2 \cdot \partial X(0) \hat k_2 \cdot \bar \partial X(0)]^{1 + \frac{\alpha' s_{12}}4} e^{i k \cdot X(0)} .
\end{align}
\label{poms}
\end{subequations}
These are the pomerons for two gauge bosons in type I, two gauge bosons in heterotic, and two gravitons in type II.  The factor $\Pi(\apr s_{12})$ is defined in Eq.~\eqref{pompi}.

The pomeron vertex operators in Eq.~\eqref{poms} display several interesting features. First, notice that they all have a power-law falloff in $z$, which comes from the exponentials common to both the bosonic and supersymmetric string. This observation is sufficient to prove that one can compute amplitudes involving these states by BCFW recursion. Also note that the tachyon pole in the gamma and $\Pi$ functions are removed by an appropriate zero in the numerator. This cancellation would not have been manifest if had we put the original vertex operators in the -1 picture, although it would be cured by vertex operators in the rest of the amplitude. 

Vertex operators and pomerons for fermionic external states work in just the same manner as we discuss in Appendix~\ref{sec:fermiPom}. In particular, all such pomerons again exhibit a power-law falloff in $z$ so that one can compute stringy amplitudes with external fermions using BCFW recursion.

\section{Subleading Terms in the Large $z$ Expansion}

In the previous section, we discussed a variety of pomeron vertex operators in various string theories and showed that any tree-level string amplitude can be computed by BCFW deformations. Our proof relied on an understanding of the leading term in the asymptotic expansion of string amplitudes in $z$. It is straightforward to extend these techniques to compute subleading terms in this asymptotic expansion. Understanding the details of this asymptotic expansion is of some intrinsic interest, but we will also see that the asymptotic expansion of certain stringy amplitudes is of a surprisingly similar form to certain field theory amplitudes.

Let us begin in the context of open string theories. In these theories, the $N$ point amplitudes are given by an integral over the positions of $N-3$ vertex operators. Specifically, the BCFW-deformed amplitude $\mathcal{M}(z)$ is given by Eq.~\eqref{eq:stringAmp}, which we reproduce here for convenience:
\begin{equation}
\label{eq:intamp}
\mathcal{M}(z) =  |w_{N,1} w_{N-1,1} w_{N,N-1}  | 
 \int \left( \prod_{i = 2}^{N-2} dw_i \right) \langle V_1(\hat k_1, w_1) V_2(\hat k_2, w_2)  V_3(k_3, w_3) \cdots V_N(k_N, w_N) \rangle.
\end{equation}
We will be concerned with the behavior of amplitudes in string theory at large $z$ so that the quantities  $\alpha' \hat s_{1j}, \alpha' \hat s_{2j}$, are large for $j \geq 3$. In this region, the $w_2$ integral in Eq.~\eqref{eq:intamp} is dominated by $w_2 \sim w_1$. Performing the (resummed) OPE of the vertex operators $V_1 V_2$ in this region generates exactly the pomeron vertex operator to leading order in $z$. Therefore, as discussed in the previous section, we can understand the large $z$ structure of string amplitudes simply by contracting the relevant pomeron operator against the other operators in the correlator.
For example, using the pomeron given in Eq.~\eqref{pomsA}, it is easy to see that the leading term in the asymptotic expansion of any amplitude involving two adjacent gauge bosons in type I string theory is given by
\begin{equation}
\label{eq:domGT}
\mathcal{M}(z) \sim (\hat \epsilon_1 \cdot \hat \epsilon_2) z^{1+\alpha' s_{12}} c,
\end{equation}
where $c$ is of order  $1 + \mathcal{O}(1/z)$. 

It is just as straightforward to compute subleading terms in the large $z$ expansion. For this purpose, we need only compute the next to leading term in the resummed OPE of the vertex operators for particles 1 and 2. The computation is most straightforward in the -1 picture. We find
\begin{multline}
V_1(0) V_2 (w) \sim \frac{ \hat \epsilon_1 \cdot \hat \epsilon_2 }{w^2} \left (1 + \frac{i}{2} w^2 \hat k_2 \cdot {\ddot X}(0) \right ) e^{i k \cdot X(0) + i w \hat k_2 \cdot \dot X(0)} w^{2 \alpha' k_1 \cdot k_2} e^{- 2 \phi (0)}\left (1 - w \dot \phi(0)\right ) \\
  - \frac{ \hat \epsilon_1 \cdot \psi(0)  \hat \epsilon_2 \cdot \psi(0) }{w} e^{i k \cdot X(0) + i w \hat k_2 \cdot \dot X(0)} w^{2 \alpha' k_1 \cdot k_2} e^{-2 \phi(0)}.
\end{multline}
We may now perform the $w$ integral. The leading order term has the same $z$ structure as in Eq.~\eqref{poms}. There are three subleading operators, which
are given by
\begin{align}
\label{eq:subleadingGT}
N_1 &= -\hat \epsilon_1  \cdot \psi(0) \,  \hat \epsilon_2 \cdot \psi(0) \Gamma(- \alpha' s_{12}) [-i \hat k_2 \cdot \dot X(0)]^{\alpha' s_{12}} e^{i k \cdot X(0)} e^{-2 \phi(0)},\\
N_2 &= \frac{i}{2} ( \hat \epsilon_1 \cdot \hat \epsilon_2 ) \,  \hat k_2 \cdot \ddot X(0) \Gamma(1 - \alpha' s_{12})[-i \hat k_2 \cdot \dot X(0)]^{-1+\alpha' s_{12}} e^{i k \cdot X(0)} e^{-2 \phi(0)}, \\
N_3 &= - (\hat \epsilon_1 \cdot \hat \epsilon_2) \dot \phi(0)  \Gamma(- \alpha' s_{12})  [-i \hat k_2 \cdot \dot X(0)]^{\alpha' s_{12}} e^{i k \cdot X(0)}  e^{-2 \phi(0)}.
\end{align} 
Notice that the polarization vector structure of $N_2$ and $N_3$ is the same as the leading order pomeron; thus, these operators contribute subleading terms in $z$ to the coefficient $c$ in Eq.~\eqref{eq:domGT}. On the other hand, $N_1$ has a different structure, as it is antisymmetric in the polarization vectors. Thus, we find the first three terms in the asymptotic expansion of the amplitude in large $z$ are
\begin{equation}
\label{eq:largezStrGB}
\mathcal{M}^{\mu \nu}(z) \sim z^{\alpha' s_{12}} \left[ \eta^{\mu \nu} (c + \cdots) z + A^{\mu \nu}  + B^{\mu \nu} \frac{1}{z} + \cdots \right],
\end{equation}
where $A^{\mu \nu}$ is an antisymmetric tensor  while $B^{\mu \nu}$ is a generic tensor, and the ellipsis indicates terms which are subdominant in $z$. The amplitude $\mathcal{M}$ is given by contracting the polarization vectors for particles 1 and 2 into $\mathcal{M}^{\mu \nu}$.
It is interesting to note that the large $z$ structure of vector amplitudes in gauge theory is given \cite{ArkaniHamed:2008yf} by 
\begin{equation}
\mathcal{M}^{\mu \nu}_{YM}(z) \sim \eta^{\mu \nu}  (c' + \cdots) z + A'{}^{\mu \nu}  + \cdots,
\end{equation}
where $A^{\prime \mu \nu}$ is also an antisymmetric tensor.
 
As another example of the applications of the pomeron we can compute the large $z$ behavior of graviton amplitudes in type II string theory. Closed string amplitudes involve integrating the position of vertex operators over the entire complex plane:
\begin{multline}
\label{eq:intampClosed}
\mathcal{M}(z) =  |w_{N,1} w_{N-1,1} w_{N,N-1}  |^2 
 \int \left( \prod_{i = 2}^{N-2} d^2w_i \right) \langle V_1(\hat k_1, w_1, \bar w_1) V_2(\hat k_2, w_2, \bar w_2)  V_3(k_3, w_3, \bar w_3) \\
 \cdots V_N(k_N, w_N, \bar w_N) \rangle.
\end{multline} 
At large $z$, the $w_2$ integrals are again dominated by the region $w_2 \sim 1/z$. Thus, the pomeron vertex operators capture exactly the leading behavior of these amplitudes in the large $z$ region, and subleading terms can be computed by calculating corrections to the resummed OPE. 

In the case at hand the work involved is simplified in the spirit of the KLT relations \cite{Kawai:1985xq}. Because the Hilbert space of the closed string has the factorized form of two copies of the open string Hilbert space, the graviton vertex operator is essentially two copies of the gauge boson vertex operators shown in Eq.~\eqref{vops}. For example, in the (-1,-1) picture, the graviton vertex operator in type II string theory is given by
\begin{equation}
V_{(-1,-1)}(w. \bar w) = \epsilon_{1 \mu \nu} \psi^\mu(w) \tilde \psi^\nu(\bar w) e^{i k \cdot X(w, \bar w)} e^{- \phi(w) -\tilde \phi (\bar w)}.
\end{equation}
This observation leads to a simplification: in the computation of any pomeron vertex operator, the contractions in the closed string case are naturally the product of two copies of the contractions involved in an open string pomeron. Moreover, it is easy to isolate the $z$ dependence of a pomeron vertex operator by a simple change of variable. The Jacobian in the closed string case is simply the square of the Jacobian in the corresponding open string case. Thus the power of $z$ present in the closed string case is twice the power in the open string case, up to the usual replacement of $\alpha'_{\mathrm closed} = \alpha'_{\mathrm open}/4$. Finally, the integrals over vertex operator positions in the closed and open string cases are different, but this affects neither the powers of $z$ nor the Lorentz structure of the polarization contractions. 
Therefore, by simply squaring the type I vector result in Eq.~\eqref{eq:largezStrGB}, we deduce that the asymptotic series for all-graviton amplitudes in type II string theory is given by
\begin{multline}
\mathcal{M}_{\mu \tilde \mu \nu \tilde \nu} = z^{\apr s_{12}/2} \left[  \eta_{\mu \nu} \eta_{\tilde \mu \tilde \nu}  (c + \cdots) z^2 + (\eta_{\mu \nu} \tilde A_{\tilde \mu \tilde \nu} + A_{\mu \nu} \eta_{\tilde \mu \tilde \nu} )  z  \right. \\
\left. +(  A_{\mu \nu \t \mu \t \nu} + \eta_{\mu \nu} \t B_{\t \mu \t \nu} +  B_{ \mu \nu} \eta_{\t \mu \t \nu} ) +  C_{\mu \nu \t \mu \t \nu} \frac{1}{z} + \cdots \right],
\end{multline}
where $A$ and $\tilde A$ are antisymmetric tensors while $B$ and $\tilde B$ are generic tensors. We also find that $C_{\mu \nu \t \mu \t \nu}$ is the sum of terms which are antisymmetric in $\mu \nu$ and in $\t \mu \t \nu$. It is remarkable that the large $z$ behavior of graviton amplitudes in general relativity is given by exactly the same formula \cite{ArkaniHamed:2008yf} without the overall factor of $z^{\apr s_{12} / 2}$, with exactly the same symmetry properties of all the tensor coefficients.

Thus we see that there appears to be a structural similarity in the asymptotic expansions of string and field theory amplitudes. This is unexpected because, of course, string theory at asymptotically large momenta is not expected to be related to field theory. In the next section, we will describe why we believe this behavior should continue at all orders in $z$ in a particular limit, at least for type I string theory and $\cN=4$ field theory.

\section{A Conjecture}

In the last section, we observed several times that stringy asymptotic structures are similar to field theory series with the same external states. It is natural to wonder if there is some limit of a string amplitude at large momentum in which the large momentum field theory limit is reproduced. Notice that this is not the usual $\alpha' \rightarrow 0$ limit because we are interested in the pomeron region of string amplitudes: therefore, we require that $\alpha' \hat s_{ij}$ is large for any deformed kinematic invariants $\hat s_{ij}$. For the purposes of this article, we will focus on the simple case of type I amplitudes with vector boson external states (and their superpartners). As we will discuss in this section, we find evidence for the conjecture that there is a large-momentum limit in which type I and $\mathcal{N} = 4$ Yang-Mills MHV superamplitudes agree.

Let us now consider the kinematic region in which we can hope this new relationship between string and field theory amplitudes can hold. Our use of the pomeron vertex operator is justified when $\alpha' \hat s_{ij}$ is large for all kinematic invariants $\hat s_{ij}$ which contain a $z$ dependence due to the BCFW deformation.  Thus it is consistent to go to a region where all other kinematic invariants are small. For example, we can take $\sqrt {\alpha' } p_i \sim \mathcal{O}(\epsilon)$ for all $i$ where $\epsilon \ll 1$ but $\sqrt {\alpha' } q \sim \mathcal{O}(\epsilon^{-1})$ so that $q \cdot p_i$ is of order 1.  We will then take $z$ to be large. In this region, $\alpha' s_{ij}  \ll 1$ while $\alpha' \hat s_{ij} \sim z$ so, for large $z$, the pomeron approximation is valid. We will refer to this particular limit of parameter space as the {\it eikonal Regge}  (ER) region.  Physically, this corresponds to a regime where one subset of momenta is much greater than the string scale while another subset is negligible compared to the string scale. Throughout this article we have considered only adjacent BCFW shifts. Correspondingly, we can consider an adjacent ER region where neighbouring particles $i, i+1$ carry large momentum. We shall restrict our attention to this adjacent ER region from now on.

We conjecture that in the adjacent eikonal Regge regime, massless MHV superamplitudes in type I string theory and $\mathcal{N} = 4$ super Yang-Mills theory are identical. In particular, our conjecture is that if two adjacent momenta $k_1$ and $k_2$ are BCFW-deformed so that $\hat s_{1i}$ and $\hat s_{2j}$, are large for $i, j \geq 3$ while all other kinematic invariants are small, then the type I amplitudes reduce to field theory amplitudes. Our evidence for this conjecture, in addition to the suggestions of the asymtotic series, is the following:
\begin{itemize}

\item Explicit proof for the four and five particle cases.

\item Demonstration that two particle factorization channels have the property that in the ER region all higher string states are suppressed for any number of particles.

\item Finally, we will show that the Berkovits-Maldacena (BM)  \cite{Berkovits:2008ic} prescription for computing string MHV superamplitudes implies our conjecture. However, the
BM expression has not yet been proven in the literature.

\end{itemize}
There may be a still stronger relation between string and field theory amplitudes--for example, perhaps $\mathrm{N^k}$MHV string and field theory amplitudes agree in the ER region, or perhaps the relationship holds for other string theory and field theory pairs. In this article, we focus only on the type I / gauge theory case. However, we cannot resist remarking that type II amplitudes with graviton external states reduce to field theory graviton amplitudes in the ER region at four and five points. This is easily seen at four points using the KLT relations; at five points it is convenient to use the BCJ~\cite{Bern:2008qj} inspired string relationship discovered by Bjerrum-Bohr, Damgaard and Vanhove~\cite{BjerrumBohr:2009rd} and by Stieberger~\cite{Stieberger:2009hq}. 

On the surface of it, the equivalence between the string and field theory amplitudes in the ER limit is quite surprising, since we are not simply taking the $\apr \rightarrow 0$ limit. Indeed, we are taking some momenta much larger than the string scale, so one would naively expect contributions from massive internal string states. Neither is this some kind of soft limit since the quantities $\alpha' \hat s_{ij}$ are large rather than small. In other words, there is plenty of energy to put more string states on-shell. Let us now discuss the evidence for our conjecture.

\subsection{Explicit Demonstrations}

Our conjecture is trivial at the level of the three point function, so we begin by examining the four point amplitude in type I string theory. For ease of presentation we will express amplitudes using the spinor-helicity formalism; this assumption can be justified by supposing that all particles are propagating in a four dimensional subspace with all polarization vectors lying in the same subspace. For our conventions, see Appendix A. In this notation, the string theory four gauge boson amplitude is 
\begin{equation}
\mathcal{A}(1^- 2^- 3^+ 4^+) = i \frac{\sh{12}^4}{\sh{12} \sh{23} \sh{34} \sh{41}} \frac{\Gamma(1 - \alpha' s_{12}) \Gamma(1- \alpha' s_{23})}{\Gamma(1 - \alpha' s_{12} - \alpha' s_{23})}.
\end{equation}
It is straightforward to see that the string theory form factor involving the $\Gamma$ functions is unity if we complex deform $k_1$ and $k_2$, since then $\alpha' s_{12}$ may be taken to be small while $\alpha' s_{23}$ is large. The remaining spinor helicity factor is exactly the gauge theory answer. Thus, our conjecture passes its first non-trivial test. It is also straightforward to show that the five particle amplitude has the desired properties using the explicit results for the MHV five point amplitude obtained by Stieberger and Taylor~\cite{Stieberger:2006te}. We describe this calculation in Appendix~\ref{sec:app5pt}.

\subsection{The Two-Particle Factorization Channel}

One of the reasons for the simplicity of the MHV amplitudes in field theory is that there are only two particle factorization channels present in the amplitude. In string theory, we no longer expect this to be the case, since there are massive resonances. Nevertheless, two particle factorization channels are straightforward to analyze. We will now consider the two particle factorization channels of an $n$ point gauge boson amplitude in type I string theory and show that the higher string modes in these factorization channels have the property that they are suppressed by a small factor in the ER region.

To simplify the calculation, we will use the usual SUSY Ward identities. The identities for $\cN = 4$ theory are expected to hold in the full type I string theory because they are purely kinematical in nature; this was carefully checked in~\cite{Stieberger:2007jv}. These identities imply the well-known identity
\begin{equation}
{\cal A}_n(1^-, 2^-, 3^+, \ldots , n^+) = \frac{\sh{12}^2}{\sh{13}^2} {\cal A}_n ( 1^-, \bar 2^0 , 3^0, 4^+, \ldots, n^+)
\end{equation}
where the states $\bar 2^0, 3^0$ are scalar antiparticles and particles. From the point of view of the 10 dimensional string theory, these scalar states are merely vectors polarized in the extra six dimensions. Now we can straightforwardly compute the OPEs of the vertex operators for these scalar states with the other particles, which are simple because many of the scalar products vanish. We shall write the scalar vertex operators as
\begin{align}
V_{-1} (w) &= e^{- \phi(w)} \Psi(w) e^{i k \cdot X(w)}, \quad V_0(w) = \frac{1}{\sqrt 2\alpha'} \left( i \dot Z(w) + 2\alpha' k \cdot \psi(w) \Psi(w) \right) e^{i k \cdot X(w)}, \nonumber\\
\bar V_{-1} (w) &= e^{- \phi(w)} \bar \Psi(w) e^{i k \cdot X(w)}, \quad  \bar V_0(w) = \frac{1}{\sqrt 2\alpha'} \left( i \dot {\bar Z}(w) + 2\alpha' k \cdot \psi(w) \bar \Psi(w) \right) e^{i k \cdot X(w)},\nonumber
\end{align}
where the non-trivial OPEs of $\Psi$ and $Z$ are
\begin{equation}
\Psi(z) \bar \Psi(0) \sim \frac{1}{z}, \quad \dot Z(z) \dot{\bar Z}(0) \sim -\frac{2 \alpha'}{z^2}.
\end{equation}
We shall take particles 1 and 2 to have the large momentum and consider the (23) factorization channel. Notice that in this factorization channel a higher string state is kinematically allowed. Choosing the vertex operators for particles 1 and 3 to be in the -1 picture, the correlator of interest is
\begin{multline}
{\cal C} = w_\infty^2 \langle [ i \dot{\bar Z}(0) + 2 \alpha' \hat k_2 \dot \psi(0) \bar \Psi(0)] e^{i \hat k_2 \cdot X(0)} e^{- \phi(w)} \Psi(w) e^{i k_3 \cdot X(w)} V_{0, 4}(w_4)\\
 \cdots
V_{0, N}(1) e^{-\phi(w_\infty)} \hat \epsilon_1 \cdot \psi(w_\infty) e^{i \hat k_1 \cdot X(w_\infty) }\rangle 
\end{multline}
where $V_{0, i}$ is a 0 picture vertex operator for the $i$th particle. We shall take $w_\infty \rightarrow \infty$ and choose the gauge $\epsilon_i \cdot k_3 = 0$ for $i = 4, \ldots, N$. Then we find 
\begin{equation}
{\cal C} = \frac{2 \alpha' w_\infty}{w} \langle \hat k_2 \cdot \psi(0) e^{i \hat k_2 \cdot X(0)} e^{i k_3 \cdot X(w)}V_{0, 4}(w_4) \cdots V_{0, N}(1) \hat \epsilon_1 \cdot \psi(w_\infty) \rangle.
\end{equation}
We can now see the structure of the two particle factorization channel. The singularities in this channel arise from the region where $w$ is small. In this region, we can perform the OPE of the operators at $0$ and $w$; since this series organizes all the $w$ dependence into Wilson coefficients we can perform the $w$ integral. The OPE is
\begin{equation}
\frac{1}{w}  e^{i \hat k_2 \cdot X(0)} e^{i k_3 \cdot X(w)} = w^{2 \alpha' \hat k_2 \cdot k_3 -1} e^{i (\hat k_2 + k_3) \cdot X(0)} \left( 1 + i w k_3 \cdot \dot X(0) + \mathcal{O}(w^2) \right)
\end{equation}
Performing the $w$ integral we see that the leading term in this OPE corresponds to a pole $1 / \hat k_2 \cdot k_3$ while the higher terms lead to the massive string poles. However, contracting the operator $\dot X(0)$ into the other vertex operators leads to factors $\alpha' k_3 \cdot k_j$ which are small. (Notice that in the channel there are no factors $1 / k_3 \cdot k_j$ because we have already factorized the leg for particle 3). We conclude that in the ER region, the higher string poles in two particle factorization channels are negligible.

\subsection{Relationship to a Conjecture of Berkovits and Maldacena}

In \cite{Berkovits:2008ic}, Berkovits and Maldacena (BM) conjecture a general form for MHV type I string amplitudes. In this section, we describe how our conjecture follows directly from theirs.

The BM conjecture is that the MHV type I superamplitude is given by 
\begin{equation}
{\cal A}(1\cdots n) = \delta^4 \left(\sum_i p_i\right) \delta^8\left(\sum_i q_i\right) \tilde {\cal A}(1\cdots n)
\end{equation}
where $p_i$ is the momentum of the $i$th particle while $q_i^{\alpha A} = \lambda_i^\alpha \eta^A$ is the supermomentum of the $i$th particle, defined in terms of four fermionic variables $\eta^A$, and
\beq
{\cal \tilde A}(1\cdots n) = \frac{1}{\sh{12} \sh{23} \sh{31}}\left \langle \left (\prod_{i=1}^3e^{i k_i X(w_i)} \right ) \left ( \prod_{j=4}^N \int^{w_1}_{w_{j-1}} dw_j \, V_{0,j} (w_j) \right ) \right \rangle,
\label{mbconj}
\eeq
where $V_{0,j}$ is the 0-picture vertex operator for the $j^{\rm th}$ gauge boson, as given in Eq.~(\ref{vops}). The first three $w_i$ are particular points, just as in any other disk amplitude. A convenient choice is to take $w_1 \rightarrow \infty, w_2=0, w_3 = 1$. Additionally, we will perform our BCFW shift on particles 1 and 2 as $\ket{\hat 1}=  \ket{1} + z \ket{2}$, $\sqet{\hat 2}  = \sqet{2} - z \sqet{1}$. 
Now, choosing the gauge for all particles $4, \ldots, N$ to be proportional to $\ket{2}$ we can compute the contractions of the operators involving $z$ to find 
\beq
{\cal \tilde A}(1\cdots n) = \frac{1}{\sh{12} \sh{23} \sh{31}}\left \langle e^{i k_3 X(1)} \left ( \prod_{j=4}^N \int^{\infty}_{w_{j-1}} dw_j \, w_j^{2 \apr k_2 \cdot k_i} V_{0,j} (w_j) \right ) \right \rangle.
\eeq
In this form we see that all dependence of the shift is in the simple factor $\prod_{i=4}^N |w_i|^{2 \apr k_2 \cdot k_i}$. 
The calculation that verifies our conjecture is now very similar to the one carried out in Appendix A.3 of \cite{Berkovits:2008ic}, in which BM show that the $\apr \rightarrow 0$ limit of Eq.~\eqref{mbconj} reproduces the field theory answer. In fact, BM show that the amplitude localizes around $w=1$, so the shifted factor is unity and does not contribute. The rest of the calculation then proceeds exactly as in BM, and we recover the $\cN =4$ result.

\section{The Pomeron and the Field Theoretic Large $z$ Expansion}

Given our conjecture on the relationship between field and string theory amplitudes, it is natural to investigate what the pomeron operators compute when the ER limit is taken. Since we know there is some relationship between string and field theory amplitudes in the ER region at four and five points, we can explore the interplay of the pomeron technique and field theory results with confidence for these amplitudes. In this section, we will focus on the simplest case of the four point amplitude, deferring our five point calculations to Appendix~\ref{sec:app5pt}. We will begin with a brief review of \cite{ArkaniHamed:2008yf} to understand the expansion of gauge theory and gravity amplitudes at large $z$. Armed with this reminder of the field theory structures, we will show how the string pomeron and subleading operators map to field theory objects in the ER region. We will then outline some explicit computations of terms in the asymptotic stringy amplitude and take the ER limit to reproduce field theory expressions.

\subsection{Review of Large $z$ Structures in Field Theory}

Individual Feynman diagrams of tree-level gauge and gravity amplitudes naively grow with energy because their associated interactions are derivatively coupled. Thus it is surprising that summing these diagrams can yield an on-shell amplitude which actually vanishes at large momenta. As we have discussed, this vanishing is crucial for the validity of the BCFW recursion relations, which require the absence of a pole at $z\rightarrow \infty$.

In \cite{ArkaniHamed:2008yf}, Arkani-Hamed and Kaplan provide a systematic description of this surprisingly convergent behavior by considering gauge and gravity amplitudes at large complex momenta. They show that the external legs which have been complex deformed can be interpreted as a hard particle propagating through a soft background corresponding to the remaining external legs. Thus, in the case of gauge theory, one can compute the large $z$ structure of amplitudes using the background field method.  Expanding around a background gauge field configuration, they obtain the Lagrangian
\begin{equation}
\mathcal{L} = - \quarter \eta^{ab} D_\mu a_a D^\mu a_b + \frac{i}{2} \Tr [a_a, a_b] F^{ab}
\end{equation}
where $F^{ab}$ is a background field containing the soft particles.  Here the indices $a, b$ are really the same as the $\mu,\nu$ Lorentz indices except they have been relabeled to emphasize what the authors of \cite{ArkaniHamed:2008yf} call a ``spin" Lorentz symmetry.  In particular, in the large $z$ limit, the term proportional to $\eta_{ab}$ dominates, and there is an enhanced Lorentz symmetry which acts on the $a,b$ indices alone.  The background field strength $F_{ab}$ explicitly breaks this enhanced symmetry at one lower order in $z$. Thus, gauge theory amplitudes at large $z$ are of the form
\beq
\epsilon_i^\mu \mathcal{A}_{\mu \nu} \epsilon_j^\nu,
\eeq
where $\epsilon_{i,j}$ are the polarization vectors of the shifted particles $i$ and $j$, and 
\beq
{\cal A}_{\mu \nu} = (c z  + \cdots ) \eta_{\mu \nu} + A_{\mu \nu} + \frac{1}{z} B_{\mu \nu}.
\eeq
where $A_{\mu \nu}$ and $B_{\mu \nu}$ are functions of the background fields, with $A_{\mu \nu}$ antisymmetric since the background field is antisymmetric. Similarly, gravity amplitudes can be calculated from a Lagrangian
\begin{equation}
L = \sqrt{-g} \left[ \quarter g^{\mu \nu}\eta^{ab} \eta^{\t a \t b} D_\mu h_{a \t a} D_\nu h_{b \t b} - \half h_{a \t a} h_{b \t b} R^{a b \t a \t b} \right],
\end{equation}
where $R^{a b \t a \t b}$ is the background curvature associated with the metric $g$ containing the soft particles and $D$ is a covariant derivative acting on the vielbein indices of the fluctuation $h$. At leading order in $z$ there is an enhanced spin Lorentz symmetry on the vielbein indices $a, b$ and also separately on the vielbein indices $\t a, \t b$. This symmetry is broken by the spin connection and by the background curvature, leading to the asymptotic form
\begin{equation}
\mathcal{M}_{\mu \tilde \mu \nu \tilde \nu} = c z^2  \eta_{\mu \nu} \eta_{\tilde \mu \tilde \nu} + z ( \eta_{\mu \nu} \tilde A_{\tilde \mu \tilde \nu} + A_{\mu \nu}  \eta_{\tilde \mu \tilde \nu} ) + A_{\mu \nu \t \mu \t \nu} + \eta_{\mu \nu} \t B_{\t \mu \t \nu} +  B_{ \mu \nu} \eta_{\t \mu \t \nu} + \frac{1}{z} C_{\mu \nu \tilde \mu \tilde \nu} + \cdots 
\label{gravamp}
\end{equation}
which is dotted into graviton polarizations in order to get the full amplitude.  
Note that $A_{\mu \nu}$ is again antisymmetric, and $A_{\mu \nu \t \mu \t \nu}$ is antisymmetric under exchange of $(\mu \nu)$ and $(\t \mu \t \nu)$. $B_{ \mu \nu}$ has no particular symmetry properties. Meanwhile, $C$ is a sum of terms antisymmetric in $(\mu \nu)$ and $(\tilde \mu \tilde \nu)$. As noted in \cite{ArkaniHamed:2008yf}, this gravitational asymptotic expansion is structurally the square of the field theory asymptotic expansion; the origin of this relationship is obscure in the field theory presentation. In both the gravity and gauge theory amplitudes, the polarizations may introduce extra powers of $z$, as we will see.

\subsection{The Pomeron and Spin Symmetry}

As we have seen, field theory asymptotic series are controlled by sources of spin symmetry violation by a background field. It is interesting to see how string theory computes the same objects. In fact, at every order in $z$, we can correlate the string theory expressions with their field theory counterparts. At leading order in large $z$, we have seen in the field theory case that both gauge and gravity amplitudes display an enhanced spin symmetry. In string theory, the leading order term in the $z$ expansion is controlled by the pomerons, given in type I for gauge amplitudes and in type II for graviton amplitudes by
\begin{align}
\eta_{\mu \nu} (1 + \apr s_{12}) \Gamma(-1 - \alpha' s_{12}) [-i \hat k_2 \cdot \dot X(0)]^{1+ \alpha' s_{12}} e^{i k \cdot X(0)} \nonumber \\
\eta_{\mu \nu} \eta_{\tilde \mu \tilde \nu}\left  (1 + \frac{\apr s_{12}}{4} \right )  \Pi (\apr s_{12})  [ \hat k_2 \cdot \partial X(0) \hat k_2 \cdot \bar \partial X(0)]^{1 + \frac{\alpha' s_{12}}2} e^{i k \cdot X(0)} \nonumber
\end{align}
respectively. Note that these operators are proportional to exactly the correct metric tensors to reproduce the enhanced spin symmetry of the field theory result. Thus, this enhanced symmetry in string theory is simply a consequence of the fact that the leading singularity in the OPEs of the vertex operators is proportional to $\eta_{\mu \nu}$.

Now let us examine these asymptotic expansions at subleading order. In the gauge theory case, we encounter the antisymmetric background field $F^{ab}$ in addition to subleading terms which preserve the spin symmetry. In type I string theory, various operators contribute at next to leading order. These operators have two kinds of Lorentz structure: they are either proportional to metric tensors (preserving the spin symmetry) or they involve an antisymmetric tensor. For example, we have already seen the subleading operator which violates the enhanced spin symmetry in the type I string (with ghost number -2); it is
\begin{equation}
\psi^\mu(0) \psi^\nu(0) \Gamma(- \alpha' s_{12}) [-i \hat k_2 \cdot \dot X(0)]^{\alpha' s_{12}} e^{i k \cdot X(0)} e^{-2 \phi(0)}.
\nonumber
\end{equation}
Of course, this operator is antisymmetric on account of the anticommutativity of $\psi$.

The comparison between the gravity amplitude and the string result is completely analogous. However, the string theoretic computation makes the KLT relation essentially manifest. So it comes as no surprise in the string case that the graviton asymptotic series is the square of the gauge boson series; this is very obscure in the field theoretic computation.

\subsection{Computational Examples}

To demonstrate how this formalism works in simple examples, we will explore some four point amplitudes in this section. Our aim is to illuminate the formal development, and also to explore the relationship between the large $z$ structure of string and field theoretic amplitudes in the context of the pomeron expansion. For simplicity we will work in the context of bosonic string theory; as before, it is also convenient to work in four dimensions. The advantage of working in four dimensions is the availability of the simple four dimensional spinor-helicity method with its compact formulae. So we imagine that all of the polarization vectors and momenta of the particles scattering happen to lie in a four dimensional subspace of a larger spacetime. Our spinor conventions are presented in Appendix~\ref{sec:spinorConv}. 

We begin with an investigation of scattering amplitudes involving four gauge bosons. In field theory, it is a well-known fact that the asymptotic behavior of the four particle Yang-Mills amplitude depends on the the helicities of the particles which are shifted. Therefore we will consider here the leading term in the asymptotic expansion for both a good and a bad shift.\footnote{A {\it good shift} is one for which the amplitude vanishes at large $z$, whereas for a {\it bad shift} the amplitude diverges.} In the case of a bad shift, we consider the amplitude $\mathcal{A}(1^- 2^+ 3^+ 4^-)$, where the superscript indicates the helicities of the gauge bosons. The shift of interest is
\begin{align}
\label{eq:badshift}
\ket{\hat 1} &=  \ket 1 + z \ket 2, \\
\sqet{\hat 2} &= \sqet 2 - z \sqet 1.
\end{align}
We choose gauges so that the polarization vectors are
\begin{align}
\hpol 1 - &= - \frac{\sqet 2 \bra{\hat 1}}{\sqrt 2 [12]}, \quad \hpol 2 + = \frac{\ket 1 \sqra{\hat 2}}{\sqrt 2 \sh{21}} \\
\pol 3 + &= \frac{\ket 2 \sqra{\hat 3}}{\sqrt 2 \sh{32}}, \quad \pol 4 - = - \frac{\sqet{\h 2} \bra{4}}{\sqrt 2 [4 \hat 2]}.
\end{align}
At leading order in the pomeron expansion, the computation reduces to calculating the expectation value of the pomeron vertex operator and two gauge boson vertex operators. Since there are no more worldsheet integrations to be performed the calculation is very straightforward. We find that
\begin{equation}
\label{eq:strBadShift}
\mathcal{A}(1^- 2^+ 3^+ 4^-) = (2 \alpha')^2 z^2 \Gamma(- \alpha' s_{12}) [\pol 3 + \cdot \pol 4 - + 2 \alpha' \pol 3 + \cdot k \; \pol 4 - \cdot p_3] (2 \alpha' \h p_2 \cdot p_3)^{1 + \alpha' s_{12}}.
\end{equation}
Now we can consider taking the eikonal Regge limit $\alpha' s_{12} \rightarrow 0$. In the ER region, the leading term of the string amplitude Eq.~\eqref{eq:strBadShift} is
\begin{equation}
\mathcal{A}(1^- 2^+ 3^+ 4^-) \rightarrow - (2 \alpha')^2 z^2  \frac{2 \h p_2 \cdot p_3 \pol 3 + \cdot \pol 4 -}{s_{12}} = (2 \alpha')^2 z^3 \frac{\sh{42}^3}{\sh{12}\sh{23}\sh{34}}.
\end{equation}
which agrees with the corresponding field theory amplitude when we correctly normalize the string amplitude.

Another interesting example at four points is given by the good shift. In this case we will consider the same amplitude $\mathcal{A}(1^- 2^+ 3^+ 4^-)$, but with the shift
\begin{align}
\label{eq:goodshift}
\sqet{\hat 1} &=  \sqet 1 + z \sqet 2, \\
\ket{\hat 2} &= \ket 2 - z \ket 1.
\end{align}
In terms of momenta, the shift is $\h p_1 = p_1 + z q, \;\; \h p_2 = p_2 -z q$ where $q = \frac{1}{2} \sqet 2 \bra 1 $.
In this case the large $z$ behavior of the field theory amplitude is
\begin{equation}
\mathcal{A}_{ft}(1^- 2^- 3^+ 4^+) = g^2 \frac{\sh{4 \h 1}^3}{\sh{12}\sh{23}\sh{34}} \sim g^2  \frac{\sh{1 4}^3}{z \sh{12}\sh{23}\sh{34}}.
\end{equation}
We would like to reproduce this behavior using our operator methods. One way to compute the large $z$ behavior of the string amplitude would be to directly compute out the various operators contributing in the pomeron expansion at order $1/z$ and above. This would be a tedious calculation since there must be a number of cancellations amongst these operators so that the leading term in $z$ is of order $z^{-1 + \mathcal{O}(\alpha' s_{12})}$. A simpler way to do the calculation is to follow the method described by Arkani-Hamed and Kaplan in~\cite{ArkaniHamed:2008yf}. We choose gauges for the external particles so that
\begin{equation}
\hpol 1 - = - \frac{\sqet 2 \bra{\h 1}}{\sqrt 2 [12]} = \pol 1 -, \quad \hpol 2 + = \frac{\ket 1 \sqra{\h 2}}{\sqrt 2 \sh{21}} = \pol 2 +,
\end{equation}
and notice that $\hpol 1 - = - \sqrt 2 q / [12]$. Since we are labeling the polarization states of the external massless bosons by polarization vectors, the Ward identity must hold to remove the unphysical polarization states. Therefore the amplitude must vanish when we replace the polarization vector of particle 1 by its momentum $\h p_1$. It follows that
\begin{equation}
q_\mu \mathcal{A}^{\mu \nu} \epsilon_{2 \nu}^+ = - \frac{1}{z} p_{1 \mu} \mathcal{A}^{\mu \nu} \epsilon_{2 \nu}^+.
\end{equation}
Furthermore, since $p_1 \cdot \pol 2 + = 0$ we see from inspecting the large $z$ structure of the stringy gauge boson amplitude given in Eq.~\eqref{eq:largezStrGB} that the leading term in the large $z$ expansion of the amplitude can be calculated simply from the antisymmetric subleading pomeron operator which is explicitly given by
\begin{equation}
N^{\mu \nu}_1 = 2 i \alpha' [k^\mu \dot X^\nu(0) - k^\nu \dot X^\mu(0)] e^{i k \cdot X(0)} [-i \h p_2 \cdot \dot X(0)]^{\alpha' s_{12}} \Gamma(-\alpha's_{12}).
\end{equation}
Performing the contractions, we find that
\begin{equation}
\mathcal{A}_{\mu \nu} = -(2 \alpha')^4 \Gamma(2 - \alpha' 2_{12}) ( k_\mu p_{3 \nu} - k_\nu p_{3\mu}) \pol 3 + \cdot \h p_2 \; \pol 4 - \cdot \h p_2 (2 \alpha' \h p_2 \cdot p_3)^{\alpha' s_{12} -2}
\end{equation}
where the polarization vectors of particles 1 and 2 are to be contracted into the $\mu$ and $\nu$ indices. Once more, it is interesting to consider the small $\alpha' s_{ij}$ limit. The result is
\begin{equation}
\mathcal{A}_{\mu \nu} = -(2 \alpha')^2 ( k_\mu p_{3 \nu} - k_\nu p_{3\mu}) \frac{ \pol 3 + \cdot \h p_2 \; \pol 4 - \cdot \h p_2}{ (\h p_2 \cdot p_3)^{2}}.
\end{equation}
Contracting in the external polarization vectors and evaluating the scalar products we find that
\begin{equation}
\mathcal{A}(1^- 2^+ 3^+ 4^-) = (2 \alpha')^2 \frac{\sh{1 4}^3}{z \sh{12}\sh{23}\sh{34}},
\end{equation}
as expected.

We present another example of the use of pomeron technology at five point order in Appendix~\ref{sec:app5pt}.

\section{Internal Recursion Relations for Tachyon Amplitudes}

We have now seen that the BCFW recursion relations hold in general for string amplitudes. However, there is a disadvantage to computing string amplitudes via BCFW, which is that the sum over intermediate states runs over an infinite set of poles.  This occurs because a string propagator describes the exchange of an infinite ladder of modes with different masses and quantum numbers.  Consequently, when BCFW is applied, for example, to an amplitude with only tachyonic external legs, the resulting lower point amplitudes will invariably include external states of higher spin.

In this section, we will explore this issue in the simplest laboratory setting: color-ordered tachyon scattering amplitudes in open bosonic string theory. We will see that for these amplitudes, the sum over intermediate states can be reexpressed using a simple new recursive relationship. In fact, it is possible to write the tachyon scattering amplitudes as a series of terms which are formed from purely tachyonic amplitudes.  Said another way, since the sum over the complete set of string theoretic states can be simply related to the lowest lying state, one can derive a new set of recursion relations which relate higher point tachyon amplitudes to lower point tachyon amplitudes.  We will refer to these as internal recursion relations. 

Throughout this section, we will use the Koba-Nielsen formula~\cite{KobaNielsen} for the $n$-particle scattering amplitude of tachyons, which is given by    
\be \MM_n = \frac{1}{\mathrm{vol} \; {\rm SL}(2,\mathbb{R})} \int [dy] \prod_{i>j} y_{ij}^{-(s_{ij}+2)} = \frac{1}{\mathrm{vol} \; {\rm SL}(2,\mathbb{R})} \int [dy] \prod_{i>j} y_{ij}^{-q_{ij}}
\label{eq:KobaNielsen}
\ee
where $y_{ij}=y_i-y_j$, $[dy] = \prod_{i=1}^N dy_i$, and we have defined $q_{ij} \equiv - 2 k_i \cdot k_j = s_{ij} + 2$. In this section we work in units where $\apr=1$. We have written the explicit division by the volume of the M\"obius group SL(2,$\mathbb{R}$) to emphasize the symmetry of the amplitude. 

\subsection{Example: the Five Point Tachyon Amplitude}

To preface a more general discussion, let us consider a simple example which illustrates how higher point tachyon amplitudes may be related to lower point tachyon amplitudes evaluated at shifted Mandelstam invariants.  Our approach at five points will largely mirror our proof of internal tachyon relations at $n$ points, which will be provided later.

For the purposes of explicit computation, we first gauge fix the general amplitude Eq.~\eqref{eq:KobaNielsen} by fixing the positions of three vertex operators. As usual, we fix the locations $y_1, y_4$ and $y_5$. 
The amplitude is given by
\beq
{\cal M}_5 = |y_{14}y_{15}y_{45}| \int dy_2 dy_3 \prod_{i<j}|y_{ij}|^{-q_{ij}}.
\eeq
Using M\"{o}bius transformations to set $y_1 = 0, y_4 =1, y_5=\infty$, relabeling $y_2 \equiv x$ and $y_3 \equiv y$, the five point tachyon amplitude becomes 
\beq
{\cal M}_5 = \int_0^1 dy \int_0^{y} dx \, x^{-q_{12}}(1-x)^{-q_{24}} y^{-q_{13}}(1-y)^{-q_{34}}(y-x)^{-q_{23}}.
\eeq
Next, we can show that this amplitude factors into a three point tachyon amplitude times a four point tachyon amplitude, with the appropriate momentum in the intermediate channel. We shift $k_1 \rightarrow \hat k_1 \equiv k_1 + z q$ and $k_5 \rightarrow \hat k_5 \equiv k_5 - zq$. Following BCFW, we need only find the poles in $z$; these can occur when the vertex operators for particles 1 and 2 become close, or when the vertex operators for particles 1, 2 and 3 become close. For the purpose of this discussion, we describe only the (1,2) channel; a similar analysis holds for the other factorization channel. Singularities in the region near $x=0$ result in poles in the $s_{12}$ channel\footnote{We suppress the hats of $z$ dependent Mandelstam invariants in this section.}. Expanding around $x=0$ the integrand becomes 
\beq
\sum_{n,m=0}^\infty x^{-q_{12}+n+m}  (1-y)^{-q_{34}} y^{-q_{13}-q_{23}-m} (-1)^{n+m} {-q_{24} \choose n}  {-q_{23} \choose  m} .
\eeq
where we have performed a binomial expansion. It is now trivial to perform the $x$ integral; it is
\begin{equation}
\int_0^y dx \, x^{-q_{12} + n + m} = \frac{-1}{q_{12} -n - m - 1} y^{-q_{12} + n + m +1}.
\end{equation}
However, we are only interested in the value of this integral at the pole in $z$ (recall that $q_{12}$ is a function of $z$.) This pole occurs when $q_{12}(z) -n - m -1 = 0$. Thus we may replace the integral by 
\begin{equation}
\int_0^y dx \, x^{-q_{12} + n + m} \rightarrow \frac{-1}{q_{12} -n - m - 1} = \frac{-1}{s_{12} -n - m + 1}.
\end{equation}
Finally, performing the $dy$ integral,
we find that 
\beq
{\cal M}_5 = \sum_{n,m=0}^\infty  \mathcal{M}_3 \frac{c_{n,m}}{s_{12} -n-m+1}  \mathcal{M}_4(q_{1+2,3} +m,q_{34}) + \cdots,
\eeq
where  the ellipsis indicates that another term corresponding to the $s_{45}$ factorization channel must be added; $q_{1+2,3} = q_{13} + q_{23} = -2(k_1 + k_2) \cdot k_3$; the three point amplitude $\mathcal{M}_3 =1$;  and $\mathcal{M}_4$ is the celebrated Veneziano amplitude, albeit evaluated at the indicated shifted Mandelstam invariants.  Furthermore, the numerator factor is
\beq
c_{n,m}=  (-1)^{n+m+1}{-q_{24} \choose n} {-q_{23}  \choose  m}.
\eeq

\subsection{Factorization channels}

Having determined how internal recursion relations work at five points, let us now generalize to $N$ points.  We start from the general tachyon amplitude Eq.~\eqref{eq:KobaNielsen}.  Like before, we gauge fix $y_1 =0$, $y_{N-1}=1$, and $y_N=\infty$.

In the previous section, we explicitly discussed the two particle factorization channel in a five point example. A completely analogous statement can be made generally for the $k$-particle factorization channel, which occurs when $k$ points pinch off on the world-sheet.  To see this, let us partition the external particles into a  ``left'' and a ``right'' group,
\be L&=&\{1,\ldots, k\} \\ R&=&\{k+1,\ldots,N\}
\ee 
and relabel the $y_{i}$ variables as $l_{i}$ or $r_{i}$, depending on whether $i$ is in $L$ or $R$.  So in other words,
\be y_{i} &=& \left\{
     \begin{array}{ll}
       l_{i} & \quad i \in L\\
       r_{i} & \quad i \in R \\
     \end{array}
   \right.
\ee
The $l_i$ and $r_i$ variables will ultimately become the moduli integrals for left and right tachyon amplitudes, $\mathcal{M}_L$ and $\mathcal{M}_R$. Because of our gauge fixing $y_1 = 0$ and $y_N = \infty$, it is convenient to define sets $L' = L \setminus \{1 \}$ and $R' = R \setminus \{N \}$. In this notation the Koba-Nielsen formula becomes
\be \MM_N &=& \int [dl][dr]\left(\prod_{i>j\in L} (l_i -l_j)^{-q_{ij}}\right)\left(\prod_{i>j\in R'} (r_i-r_j)^{-q_{ij}}\right)\left(\prod_{i\in L ,j\in R'} (r_{j}- l_{i} )^{-q_{ij}}\right), \nonumber
\ee
where the measure $\int [dl][dr]$ is shorthand for
\begin{equation}
\int [dl][dr] = \int_0^1 dr_{N-2}\int_0^{r_{N-2}}dr_{N-3}\cdots \int_0^{r_{k+1}}dl_k \cdots \int_0^{l_4} dl_3 \int_0^{l_3} dl_2
\end{equation}
The $k$-particle factorization channel occurs when
\be s_{12\ldots k}&=&-\left(\sum_{i =1}^k k_i\right)^2 \\
&=& -k+ \sum_{i>j}q_{ij} \\ &=&-1+n.
\ee  
We are now in a position to expose the corresponding singularity in the $N$ tachyon amplitude.  The $k$-particle factorization channel corresponds to the limit in which particles 1 through $k$ coincide on the worldsheet.  This occurs when $l_k \rightarrow 0$.  In order to parameterize this limit, we wish to define $l_k\equiv C$ as the ``pinch'' variable, and rescale the remaining variables
\be l_{i} \rightarrow C l_{i} , \ee  
for all $i\in L \setminus \{k\}$.  Notice that while we are formally rescaling $l_1$, this does nothing because this variable has been gauge fixed to zero.

Now when we then take $C \rightarrow 0$, the vertex operators corresponding to particles 1 through $k$ collapse to a single point, yielding the $k$-particle factorization channel.  Including the resulting Jacobian, the amplitude becomes
\begin{multline}
\MM_N = \int [dl][dr] dC\; C^{k-2}\left(\prod_{i>j\in L} [C (l_i-l_j)]^{-q_{ij}}\right)\left(\prod_{i>j\in R'} (r_i-r_j)^{-q_{ij}}\right)
\\
\times \left(\prod_{i\in L ,j\in R'} (r_{j}-C l_{i})^{-q_{ij}}\right) .
\end{multline}
In the above expression, $l_1=0$, $r_{N-1}=1$ and $r_{N}=\infty$ due to our original gauge fixing of $\mathcal{M}_N$.  Moreover, we can think of our rescaled ``pinch'' variable $l_k=C$ as having been gauge fixed to one, $l_k=1$, and then multiplied by $C$.  Setting $l_k$ to unity corresponds to one of the gauge fixings of $\mathcal{M}_L$.

We can pull out a factor of 
\be C^{-\left(\sum_{i,j\in L} q_{ij} -k + 2 \right)}&=& C^{-(s_{12\ldots k} +2)}
\ee
which yields the multi-particle singularity in the integration region where $C$ is small. Now, let us expand in small $C$, looking at only the relevant terms
\begin{multline} 
\label{eq:cint} \int dC \;  C^{-(s_{12\ldots k} +2)}\left(\prod_{i\in L ,j\in R'} (r_{j}-C l_{i})^{-q_{ij}}\right)  \\
= \int dC \; C^{-(s_{12\ldots k} +2)} \left( \prod_{i\in L' ,j\in R'} \left[ \sum_{n_{ij}=0}^\infty(-1)^{n_{ij}} C^{n_{ij}} \; l_{i}^{n_{ij}}\;  r_{j}^{-(q_{ij}+n_{ij})} \left(\begin{array}{c}
    -q_{ij} \\ n_{ij}
\end{array}\right) \right] \right) \left( \prod_{j \in R'} r_j^{-q_{1j}} \right). 
\end{multline}
Performing the $C$ integral near the pole in $s_{12 \ldots k}(z)$, we may write the amplitude near the $k$ particle factorization channel as
\begin{multline}
\MM_N \sim \sum_{n=0}^{\infty}\int [dl] [dr] \frac{-1}{s_{12 \ldots k} + 1 -n} 
\left( \prod_{i > j \in L} l_{ij}^{-q_{ij}} \right)
\left( \prod_{i > j \in R'} r_{ij}^{-q_{ij}} \right) \\
\left( \prod_{j \in R'} r_j^{-\sum_{i \in L} q_{ij}} \right) 
\left( \prod_{i \in L', \, j \in R'}  \sum_{\rm{part}} (-1)^{n_{ij}}
\left(\begin{array}{c}
    -q_{ij} \\ n_{ij}
\end{array}\right)
l_i^{n_{ij}} r_j^{-n_{ij}}
 \right)
\end{multline}
where 
``part'' denotes a sum over partitions of $n$ into $(N-1)(N-k-1)$ numbers $n_{ij}$ such that $\sum n_{ij} = n$ for $i\in L'$ and $j\in R'$.  Note that these manipulations have  factorized the amplitude in terms of $l$ and $r$ worldsheet coordinates corresponding to the $L$ and $R$ groups of particles separated by the factorization channel. 

\subsection{Recursion Relation}

We have systematically extracted the residue corresponding to the $s_{12\ldots k}$ factorization channel in which the amplitude is split into a $L$ and $R$ group.  With this understanding of the factorization properties of the tachyon amplitudes, and using the usual BCFW logic which allows us to reconstuct the amplitude from these factorization channels, we can see how to organize the tachyon amplitude to reveal a new recursion relation involving tachyon amplitudes only. 
%
We can write the amplitude as
\begin{multline}
\MM_N = \sum_k \sum_n \sum_{\rm part}  \frac{(-1)^{1+n}}{s_{12 \ldots k} + 1 -n} 
\prod_{ij} \left(\begin{array}{c}
    -q_{ij} \\ n_{ij}
\end{array}\right)
\int [dl][dr] 
\left(\prod_{i>j\in L} l_{ij}^{-q_{ij}}\right) \left(\prod_{i\in L'} l_{i}^{\sum_{j \in R'} n_{ij}}\right)  \\
\left(\prod_{i>j\in R'} r_{ij}^{-q_{ij}}\right) \left( \prod_{j \in R'} r_j^{-\sum_{i \in L} q_{ij}} \right)  \left(\prod_{j\in R'} r_{j}^{-\sum_{i \in L'} n_{ij}}\right) 
.
\end{multline}
The integrals over the $l_i$ and $r_j$ are completely disentangled. To understand the structure of these integrals, it is helpful to 
write the factor $l_{i}^{n_{ij}}$ more suggestively as $(l_i - 0)^{n_{ij}} = (l_i - l_1)^{n_{ij}}$.  Multiplication by all such factors is equivalent to shifting $q_{1i}\rightarrow \hat q_{1i} = q_{1i}- \sum_{j\in R'} n_{ij}$. We define $\hat q_{ij} = q_{ij}$ for $i, j \in L'$. Note that $\hat q_{1k} = q_{1k}$ since $l_k = 1$. Then the integrals over the $l_i$ can be written as a shifted tachyon amplitude,
\begin{equation}
\int [dl] \left(\prod_{i>j\in L} l_{ij}^{-\hat q_{ij}}\right) = \widehat \MM_L(\hat q_{ij}).
\end{equation}
Similarly, the integrals over the $r_j$ can be written as a right-hand tachyon amplitude at shifted kinematics. To do so, it is helpful to introduce $r_L = 0$ as the gauge-fixed position of the intermediate leg as it enters the right-hand amplitude; this leg carries all the momentum $k_L = \sum_{i \in L} k_i$ flowing into the left-hand diagram. Then the factor $r_{j}^{-(q_{ij}+ n_{ij})}$ may be written more suggestively as $(r_{j}-r_L)^{-(q_{ij}+ n_{ij})}$. We define $q_{Lj} = \sum_{i \in L} q_{ij} = -2 k_L \cdot k_j$. The variable is the analogue of $q_{1i}$ in the left-hand amplitude. We further define $\hat q_{Lj} = q_{Lj} + \sum_{i \in L'} n_{ij}$ and $\hat q_{ij} = q_{ij}$ for $i,j \in R$. We can now evaluate the integrals over the $r_j$ as
\begin{equation}
\int [dr] \left(\prod_{i>j\in R' \cup \{ r_L \}} r_{ij}^{-\hat q_{ij}}\right) = \widehat \MM_R(\hat q_{ij}).
\end{equation}
Notice that, in effect, we have found that our left-hand amplitude is gauge-fixed with particle 1 at position 0, particle $k$ at position 1 and an intermediate particle gauge-fixed at position $\infty$. In the right-hand amplitude we have found the intermediate particle to be gauge-fixed at position 0 while particles $N-1$ and $N$ inherited their gauge fixing from the original $N$ point amplitude.

These manipulations yield our final recursion relation, which involves the usual BCFW factorization of an amplitude into a left and right sub-amplitude, a sum over the mass level $n$ of the string, as well as a sum over partitions $\{n_{ij}\}$ of $n$:
\be
\label{eq:genRR}
\MM_N &=& \sum_{L,R} \sum_{n =0}^\infty \sum_{\{n_{ij}\}}\widehat\MM_L(\hat q) \frac{{\rm res}(\{n_{ij}\})}{s_{12\ldots k} +1 -n}  \widehat\MM_R(\hat q) \\
{\rm res}(\{n_{ij}\}) &=& (-1)^{n+1} \prod_{n_{ij}} \left(\begin{array}{c}
    -q_{ij} \\ n_{ij}
\end{array}\right). 
\ee
Of course, $\widehat \MM_L(\hat q)$ is a function of Mandelstam variables. For us, a convenient choice is to pick $q_{ij}$ with $i < j \in L$, and to omit $q_{1k}$. Similarly, we choose $\widehat \MM_R(\hat q)$ to be a function of $q_{ij}$ for $i<j \in R$, but omitting $q_{iN}$. To include the kinematics of the intermediate state, we additionally make $\MM_R$ a function of $q_{Lj}$ for all $j \in R$ except $N$ and $N-1$.
As stated above, the $\widehat\MM_L$ and $\widehat\MM_R$ in appearing in Eq.~\eqref{eq:genRR} are functions of integer shifted Mandelstam variables. In particular, the necessary shifts are given by
\be
q_{1i} &\rightarrow& \hat q_{1i} = q_{1i}- \sum_j n_{ij}, \, {\rm with} \,\, {i = 2,\cdots,k -1}\,\, {\rm and} \,\, {j = k+1, \cdots, N-1} \\
q_{Lj} &\rightarrow& \hat q_{Lj} = q_{Lj} + \sum_i n_{ij},   {\rm with} \,\,  {i=2,\cdots,k} \,\, {\rm and}  \,\, {j = k+1, \cdots, N-2}. 
\ee
This recursion relation is distinct from BCFW in the sense that there is no explicit sum over intermediate states---the lower point amplitudes only involve tachyons as external states, and they are evaluated at Mandelstam invariants which have been shifted by integer values.  From this point of view, the entire effect of the intermediate state sum is encapsulated by the factor $\textrm{res}(\{n_{ij}\})$. 

The partition of $n$ appearing in the recursion relation Eq.~\eqref{eq:genRR} requires some explanation.  The binomial coefficients in $\textrm{res}(\{n_{ij}\})$ come from expanding differences of vertex operator positions when one vertex operator is on the left and the other is on the right. However, if the vertex operator on the left has been fixed at zero, there will be no binomial expansion. Similarly, if the right vertex operator is at infinity, there will be no expansion. Thus, for a $k$ particle factorization channel in an $N$ point amplitude, there are $(k-1)(N-k-1)$ binomial expansions. Consequently, at mass level $n$, the $n_{ij}$ consist of partitions of $n$ into $(k-1)(N-k-1)$ integers. 

The simplification that has occurred in the sum over intermediate states can be understood in terms of the OPE of the tachyon vertex operators, which are simply given by $e^{i k \cdot X}$. This is most obvious for two particle factorization channels though the result is quite general. Let us consider a singularity in the amplitude when particles 1 and 2 join. In this region, the vertex operators for the particles are close together. The OPE is
\begin{align}
e^{i k_1 \cdot X(0)} e^{i k_2 \cdot X(w)} &= w^{2 k_1 \cdot k_2} e^{i k_1 \cdot X(0) + i k_2 \cdot X(w)} \nonumber \\
&= w^{-2 -  s_{12}} e^{i (k_1 + k_2)  \cdot X(0)} \left( 1 + w k_2 \cdot \dot X(0) + \cdots \right).
\end{align} 
Now, performing the $w$ integral, we obtain an infinite series of poles corresponding to the various masses of string states. The residues of each of the poles are simply related because of the structure of the OPE. More general amplitudes involve slightly more complicated vertex operators, but the OPEs of these vertex operators are still relatively simple objects. Therefore we expect these internal recursion relations to occur quite generally, albeit in a more complicated form.

\section{Conclusions}
\label{sec:conclusions}

In this work we have shown that all tree-level perturbative string theory amplitudes can be computed via  BCFW recursion. Our proof relied on the pomeron vertex operator technology developed in~\cite{Brower:2006ea}. We explored string amplitudes with massless external states as an asymptotic series in large $z$, and found remarkable structural similarities to amplitudes in the corresponding quantum field theory. This led us to conjecture that massless type I string amplitudes reduce to $\mathcal{N} = 4$ super-Yang Mills not only in the small $\alpha'$ limit, but also in what we term the eikonal Regge limit. This limit corresponds to taking $\alpha' \hat s_{ij}$ to be large for kinematic invariants that receive an adjacent BCFW deformation while all other independent $\alpha' s_{kl}$ are small. We have seen that our conjecture is true in several non-trivial examples and provided evidence at all points.

Nevertheless, BCFW recursion techniques applied to string theoretic amplitudes suffer from a disadvantage; in particular, since a string propagator describes an infinite ladder of states, the BCFW sum is necessarily infinite. Thus, to compute the four point tachyon amplitude in bosonic string theory, \emph{i.e.} the Veneziano amplitude, one must sum over an infinite set of three point functions describing the interaction of two tachyons and an arbitrary string state. However, an exploration of the structure of the sum appearing in the $n$ tachyon amplitude has revealed a new recursion relation which allows one to write the amplitude in terms of on-shell lower point amplitudes with only tachyonic external states. We believe that a similar structure exists in general for all string amplitudes.

Our work has lead to several new questions which we feel are worthy of further exploration.  In particular, it would be interesting to explore the possibility of additional internal recursion relations for other classes of string amplitudes beyond bosonic tachyon amplitudes. It may be that there is a way of organizing this recursion relation which makes clear that the object being summed over is the full string multiplet, analogous to the integration over the full set of states in the $\mathcal{N} = 4$ Yang-Mills theory. In particular, a new level of insight into stringy amplitudes could be achieved if one can develop a method of parameterizing the full superstring multiplet in ten dimensions. To make progress in this direction it may first be necessary to understand on-shell superspace in 10 dimensions. Progress on generalizing the four dimensional superspace methods has recently been made in~\cite{Cheung:2009dc,Dennen:2009vk,Boels:2009bv}  so this may be an achievable first step.

Of course, it would be of great interest to understand more clearly the origin of the structural similarity of the asymptotic expansions of string and field theoretic amplitudes. In this vein, a deeper study of the conjecture we made relating string amplitudes in the ER region to their effective field theory amplitude would be warranted. Since the four and five point graviton amplitudes in type II string theory reduce to field theoretic graviton amplitudes in the ER region, it may also be worth investigating the ER limit in type II string theory more thoroughly, especially in light of the KLT-like structure we have found.

Finally, our focus in this work has been on scattering perturbative string states. As observed by BPST~\cite{Brower:2006ea}, there is no obstruction to applying pomeron techniques to study D-brane scattering processes. Therefore it seems likely that one can compute scattering amplitudes for non-perturbative states using BCFW techniques, and this could lead to new insights into the physics of these nonperturbative objects.

\begin{center}
\bf{Acknowledgments}
\end{center}
\medskip
We would like to thank Nima Arkani-Hamed and Juan Maldacena for useful discussions. CC is supported by the National Science Foundation under grant PHY-0555661. DOC is supported in part by DOE grant DE-FG02-90ER40542, and by the Martin A. and Helen Chooljian Membership at the Institute for Advanced Study.  BW is supported in part by DOE grant DE-FG02-90ER40542, and by the Frank and Peggy Taplin Membership at the Institute for Advanced Study. 

\appendix

\section{Conventions}
\label{sec:spinorConv}

In this section, we review our conventions, and also state a few formulae useful for re-deriving our results.

We work with the flat metric 
\beq
\eta_{\mu \nu} = {\rm diag}(-,+, \cdots,+).
\eeq
For $\sigma$ matrices, we define
\begin{equation}
\sigma_0 = \begin{pmatrix}
1 & 0 \\
0 & 1
\end{pmatrix},
\;\; \sigma_1 = \begin{pmatrix}
0 & 1 \\
1 & 0
\end{pmatrix},
\;\; \sigma_2 = \begin{pmatrix}
0 & -i \\
i & 0
\end{pmatrix},
\;\; \sigma_3 = \begin{pmatrix}
1 & 0 \\
0 & -1
\end{pmatrix}.
\;\;
\end{equation}
The associated tilde matrices are
\begin{equation}
\t \sigma_0 =- \begin{pmatrix}
1 & 0 \\
0 & 1
\end{pmatrix},
\;\; \t \sigma_1 = \begin{pmatrix}
0 & 1 \\
1 & 0
\end{pmatrix},
\;\; \t \sigma_2 = \begin{pmatrix}
0 & -i \\
i & 0
\end{pmatrix},
\;\; \t \sigma_3 = \begin{pmatrix}
1 & 0 \\
0 & -1
\end{pmatrix}.
\;\;
\end{equation}
These satisfy the Clifford algebra
\begin{equation}
\sigma^\mu \t \sigma^\nu + \sigma^\nu \t \sigma^\mu = 2 \eta^{\mu \nu}.
\end{equation}
As usual, we take the $SU(2)$ indices of the $\sigma$ matrices to be $\sigma^\mu_{\alpha \d \alpha}$ and $\t \sigma^{\mu \d \alpha \alpha}$. We define $\epsilon$ matrices with upper and lower, dotted and undotted, indices, 
\begin{equation}
\epsilon_{\alpha \beta} = \begin{pmatrix}
0 & -1 \\
1 & 0
\end{pmatrix} , \;\;\;
\epsilon_{\d \alpha \d \beta} =  \begin{pmatrix}
0 & 1 \\
-1 & 0
\end{pmatrix}.
\end{equation}
These satisfy $\epsilon^{\d \alpha \d \beta} \epsilon^{\alpha \beta} \sigma^\mu_{\beta \d \beta} = \t \sigma^{\mu \d \alpha \alpha}$, as well as the usual Fierz relations
\begin{equation}
\sigma^\mu_{\alpha \d \alpha} \sigma_{\mu \beta \d \beta} =  2 \epsilon_{\alpha \beta} \epsilon_{\d \alpha \d \beta}.
\end{equation}

We take the relation between momenta and the corresponding spinors to be $ p \cdot \sigma_{\alpha \d \alpha} = \lambda_\alpha \t \lambda_{\d \alpha}$, which means that $p^\mu = {1 \over 2} \lambda \t \sigma^\mu \lambda = {1 \over 2} \lambda \sigma^\mu \t \lambda$. Taking another momentum $q^\mu = {1 \over 2} \zeta \sigma^\mu \t \zeta$, scalar products are given by
\begin{equation}
2 p \cdot q = \langle \lambda \zeta \rangle [\t \lambda \t \zeta],
\end{equation}
where the brackets are defined by:
\begin{equation}
\langle \lambda \zeta \rangle = \epsilon_{\alpha \beta} \lambda^\alpha \zeta^\beta , \;\;\; [ \t \lambda \t \zeta ] = \epsilon_{\d \alpha \d \beta} \t \lambda^{\d \alpha} \t \zeta^{\d \beta}.
\end{equation}
In this work, we define kinematic invariants $s_{ij} \equiv -( p_i + p_j)^2$. With this convention,
\begin{equation}
s_{ij} = \langle i j \rangle [j i ].
\end{equation}

For a particle of momentum $p$, we choose a reference momentum $q$ so the positive and negative helicity vectors are
\begin{eqnarray}
\epsilon_+^\mu &=& \frac{ \langle \lambda | \sigma^\mu | \zeta ]}{\sqrt 2 [\t \lambda \t \zeta]} \\
\epsilon_-^\mu &=& -\frac{[ \tilde \lambda | \t \sigma^\mu | \zeta \rangle}{\sqrt 2 \langle \lambda \zeta \rangle}.
\end{eqnarray}
These vectors satisfy $\epsilon_+ \cdot \epsilon_- = 1$ with all other inner products vanishing.

Finally, we note for convenience the OPE
\begin{equation}
X^\mu (w) X^\nu (z) \sim - \frac{\apr}2 \eta^{\mu \nu} \ln |w-z|^2,
\end{equation}
which means that when we restrict to the boundary of the upper half plane
\beq
X^\mu (y_1) X^\nu (y_2) \sim - 2\apr \eta^{\mu \nu} \ln y_{12}.
\eeq
Additionally, the worldsheet spinors $\psi^\mu$ satisfy
\beq
\psi^\mu(w) \psi^\nu(z) \sim \frac{\eta^{\mu \nu}}{w-z}.
\eeq

\section{Gaugino Vertex Operators and Pomerons}
\label{sec:fermiPom}

In this section, we record some pomeron vertex operations involving gauginos. 
In the type I string, the gaugino vertex operator in the $-1/2$ picture is 
\beq
V_{-1/2} =  (\apr)^{1/4} \,  u_\alpha \Theta_\alpha  e^{ik \cdot X} e^{-\phi/2},
\eeq
where $\Theta_\alpha$ is the spin field operator and $u_\alpha$ is the polarization, while $\alpha$ is a ten-dimensional Majorana-Weyl spinor index. Although any amplitude with an odd number of fermions will vanish, there is a sensible pomeron for a gaugino and gauge boson. It is most convenient to work in the -1/2 picture for the gaugino, and -1 picture for the gauge boson. Using the OPE
\beq
\left ( \Theta_\alpha (0) u_\alpha e^{-\phi(0)/2} \right ) \left (  \epsilon_\mu \psi^\mu(w) e^{-\phi(w)} \right ) \sim \frac{1}{w \sqrt 2} \epsilon_\mu \Gamma^\mu_{\alpha \beta} u_\alpha \Theta_\beta(0) e^{- \phi(0)/2 - \phi(w)}
\eeq
we find that the pomeron for a gaugino and a gauge boson in type I is
\beq
u_\alpha \Theta_\beta (0) \epsilon_\mu \Gamma^\mu_{\beta \alpha} e^{-3\phi(0)/2} \Gamma(-\apr s_{12}) \left ( - i \hat k_2 \cdot \dot X(0) \right)^{\apr s_{12}}.
\eeq

We also list for reference the pomeron for two type I (same-helicity) gauginos with polarizations $u_\alpha, v_\alpha$.
\beq
u_\alpha (C \Gamma^\mu)_{\alpha \beta} v_\beta e^{-\phi(0)} \psi_\mu(0) \Gamma(-\apr s_{12}) \left ( - i \hat k_2 \cdot \dot X(0) \right)^{\apr s_{12}}.
\eeq
These fermionic pomerons again exhibit a power-law falloff in $z$ so that amplitudes involving these external states can be computed using the BCFW recursion relations.

\section{Pomeron Technology and Five-Point Amplitudes}
\label{sec:app5pt}

In this appendix, we do two calculations with five-point MHV amplitudes. First, we demonstrate that our conjecture is valid for the type I five-point amplitude, using a result from  \cite{Stieberger:2006te}. Additionally, we check the leading behavior in $z$ for the five-point bosonic string amplitude.

\subsection{Type I at All Orders in $z$}

The five point function of gauge bosons in type I has been presented in the language of the spinor-helicity formalism by \cite{Stieberger:2006te}. The result is given by
\begin{equation}
\mathcal{A} = \left [ V(s_{ij}) + P(s_{ij}) \alpha'^2 \epsilon(1,2,3,4) \right ] \mathcal{A}_\mathrm{YM},
\end{equation}
where $\mathcal{A}_\mathrm{YM}$ is the Yang-Mills amplitude, $\epsilon(1,2,3,4) = \epsilon_{\mu \nu \rho \sigma} k_1^\mu k_2^\nu k_3^\rho k_4^\sigma$, and
\begin{equation}
V \equiv s_{23} s_{51} f_1 + \half (s_{23} s_{34} +s_{45} s_{51} -s_{12} s_{23} -s_{34} s_{45} -s_{12} s_{51}) f_2, \;\;\;
P \equiv f_2,
\end{equation}
where the functions $f_1$ and $f_2$ are given in terms of a hypergeometic function
\begin{multline}
F \left[ {n_1, n_2 \atop n_{11} , n_{12}, n_{22}} \right] = \frac{\Gamma(s_{23} + n_1 - 2 ) \Gamma( s_{15} + n_2 -1 ) \Gamma( s_{34} + n_{11} +1 ) \Gamma( s_{45} + n_{22} + 1) }{\Gamma(s_{23} + s_{34}  + n_1 +n_{11} -1) \Gamma( s_{51} +s_{45}  + n_2 +n_{22})} \\
\times {}_3F_2 \left[ { s_{23} + n_1 -2, s_{15} + n_2 - 1 , -s_{35}-n_{12} \atop s_{23} + s_{34} + n_1 + n_{11} -1 , s_{51} + s_{45} +n_2 + n_{22} };1 \right]
\end{multline}
by
\begin{equation}
f_1=F\left[ {2,1\atop 0,0,0} \right] \qquad \makebox{and} \qquad f_2= F \left[ {3,2 \atop 0,-1,0} \right].
\end{equation}
The functions $V$ and $P$ are cyclically symmetric in the particle number so without loss of generality we can consider deforming the momenta of particles $1$ and $2$. In the region $s_{12} =  s_{34} =  s_{45} = s_{35} = 0$ we find that $V = 1$ and 
\begin{equation}
P = \frac{\psi(\alpha' \hat s_{23}  + 1) - \psi (\alpha' \hat s_{51} + 1)} {\alpha' (\hat s_{23} -\hat s_{51}) },
\end{equation}
where $\psi(x) \equiv \Gamma^\prime (x)/\Gamma(x)$ is the digamma function.
Thus, the quantity $P \alpha'^2 \epsilon(1,2,3,4) \rightarrow 0$ in the ER limit so that our conjecture holds at five points in type I string theory. It would, of course, be of great interest to check the conjecture at higher points; however, beyond five point order the amplitudes can no longer be expressed in terms of hypergeometric functions so progress is more difficult.

\subsection{The Bosonic String at Leading Order in $z$}

The details of this calculation are somewhat tedious, so here we summarize the basic points. As in the four point amplitude, we  use the pomeron vertex operator. In contrast to the four-point calculation, however, we have one integral we need to evaluate. The full expression is 
\beq
\int dw \left ( -i \hat k_2 \cdot \dot X(0) \right)^{1 + \apr s_{12}} e^{i k \cdot X(0)} \epsilon_3 \cdot \dot X(w) e^{i k_3 \cdot X(w)} \epsilon_4 \cdot \dot X(w_4) e^{i k_4 \cdot X(w_4)} \epsilon_5 \cdot \dot X(w_5) e^{i k_5 \cdot X(w_5)},
\eeq
where we have omitted a prefactor of $C_{12}(z) \Gamma(-1-\apr s_{12})$. It is convenient to fix $w_4 = 1, w_5 \rightarrow \infty.$
The various contractions of polarizations and momenta will then give different powers of $w$ and $(1-w)$, some of which get contracted into part of the pomeron vertex operator. Using the OPE
\beq
\left ( -i \hat k_2 \cdot \dot X(0) \right )^n e^{i p_i \cdot X(w_i)} e^{i p_j \cdot X(w_j)} \sim (-2 \apr)^n \left [ \frac{ \hat k_2 \cdot p_i}{-w_i} +  \frac{ \hat k_2 \cdot p_j}{-w_j} \right ]^n e^{i p_i \cdot  X(w_i)} e^{i p_j \cdot  X(w_j)}.
\eeq
Since we take $w_5 \rightarrow \infty$, the only contractions that survive have powers of $w$ and $1-w$. In general, then, the integrals we must do are of the form
\beq
I(p,q) \equiv \int dw \, \left ( \frac{\hat k_2 \cdot k_3}{w} + \hat k_2 \cdot k_4\right )^{1+\apr s_{12}}  w^p (1-w)^q w^{2 \alpha' k_3 \cdot k} (1-w)^{2 \alpha' k_3 \cdot k_4},
\eeq
where the powers of $p$ and $q$ come from contractions with $\epsilon \cdot \dot X$, and the $w^{\apr k_i \cdot k_j}$ come as usual from contractions between the exponentials. The full answer is a sum of a number of different terms involving different powers of $p$ and $q$.

The terms in the five-point amplitude have coefficients of the form $(\epsilon \cdot k)^3$ or $(\epsilon \cdot \epsilon) (\epsilon \cdot k)$. In the eikonal Regge limit, we need not worry about the former because they will always be one power of $\apr$ higher than the latter. We find that the final expression is 
\beq
\cA(1^- 2^+ 3^+ 4^+ 5^- ) \sim P(z,\alpha') Q(\alpha',z),
\eeq
where
\beq
P(z,\alpha') \equiv iC_{12}(z) \Gamma(-1 -\alpha' s_{12}) (2 \alpha')^{4+\alpha' s_{12}} 
\eeq
and
\begin{eqnarray}
Q(\alpha',z) &\equiv& (\epsilon_3^+ \cdot \epsilon_5^-) (\epsilon_4^+ \cdot k) I(0,0) + 
(\epsilon_3^+ \cdot \epsilon_5^-) (\epsilon_4^+ \cdot k_3) I(0,-1) \cr &+&(\epsilon_4^+ \cdot \epsilon_5^-) (\epsilon_3^+ \cdot k) I(-1,0)-(\epsilon_4^+ \cdot \epsilon_5^-) (\epsilon_3^+ \cdot k_4) I(0,-1) 
\label{qis}
\end{eqnarray}
We find that each of the integrals in Eq.~\eqref{qis} is $\cO(1/\apr)$ as $\apr \rightarrow 0$. As a result, the full amplitude goes as $\apr^{2}$.

Since $C_{12}(z) \sim z^2$ and $Q(\apr,z) \sim z$, we reproduce the $z^3$ behavior of the bad shift. This much was nearly automatic from the beginning and is no surprise. However, we find precise agreement between the field theory amplitude and the small $\apr$ behavior of the string theory amplitude, as conjectured.

\end{document}